\documentstyle[twoside,fleqn,espcrc2,psfig,here]{article}
\newcommand{\be}{\begin{eqnarray}}
\newcommand{\ee}{\end{eqnarray}}
\newcommand{\bea}{\begin{eqnarray}}
\newcommand{\eea}{\end{eqnarray}}
\newcommand{\ba}{\begin{array}}
\newcommand{\ea}{\end{array}}

\newcommand{\no}{\nonumber}

\newcommand{\Op}{{\cal O}}

\newcommand{\lw}[1]{\smash{\lower1.6ex\hbox{#1}}}
\def\simgt{\rlap{\lower 3.5 pt\hbox{$\mathchar \sim$}}
           \raise 1pt \hbox {$>$}}
\def\simlt{\rlap{\lower 3.5 pt\hbox{$\mathchar \sim$}}
           \raise 1pt \hbox {$<$}}

\newcommand{\AmS}{{\protect\the\textfont2
  A\kern-.1667em\lower.5ex\hbox{M}\kern-.125emS}}
\title{Preliminary study of $B_B$ parameter using Lattice NRQCD}
\author{
N. Yamada\address{Department of Physics, Hiroshima University,
Higashi-Hiroshima 739-8526, Japan
}\thanks{
Presented by N.Yamada.},
S. Hashimoto\address{Fermi National Accelerator Laboratory,
P.O. Box 500, Batavia, IL 60510
}$^{\rm ,d}$, 
K-I. Ishikawa$^{\rm a}$,
H. Matsufuru\address{Institut f\"ur Theoretische Physik,
                     Universit\"at Heidelberg, Heidelberg, Germany
}\thanks{H.M. and S.T. would like to thank
the JSPS for Young Scientists for a research fellowship.},
T. Onogi$^{\rm a}$\thanks{T.O. is supported by the Grants-in-Aid
of the Ministry of Education (No. 10740125).} and
S. Tominaga\address{High Energy Accelerator Research
Organization(KEK),Tsukuba 305-0801, Japan
}$^\dag$
}
\begin{document}
\begin{abstract}
We present the preliminary result of the calculation of $B_B$ parameter
using lattice NRQCD.
The calculation is performed on a $16^3 \times 48$ quenched lattice
at $\beta=5.9$ with clover light quark.
The use of lattice NRQCD enables us to investigate the $1/m_Q$ correction
from the static limit.
The observed mass dependence is well described by the vacuum
saturation.
\end{abstract}
\maketitle
\section{Introduction}

The theoretical determination of the $B$ meson decay
constant $f_B$ and the bag parameter $B_B$ are necessary
to constrain $|V_{tb}^*V_{td}|$.
The lattice technique has been considered to be the most
reliable approach, and a number of authors have calculated
these quantities in different approaches.
In recent years, the value of $f_B$ is almost being settled
(but still within the quenched approximation) \cite{draper}.
On the other hand, $B_B$ has been calculated only under the
approximation of $m_b\rightarrow\infty$ or $a^{-1}\gg m_b$,
and there has been no systematic study of its heavy quark
mass dependence.
In this talk we study the $1/m_Q$ correction to the static
limit using the NRQCD action, which is the first step to
obtain $B_B$ with the realistic $b$-quark mass.

\section{Method}

Since the chiral symmetry is broken on the lattice, several
lattice operators contribute to $B_B$ through
\be
B_B(m_b) = \sum_{i} Z_i B_i^{\rm lat} \label{eq1},
\ee
where $B_i^{\rm lat}$ are
\be
    B_i^{\rm lat}(a^{-1})
&=& \frac{\langle \overline{B^0}|\
    \hat{\Op^l_i}(a^{-1})\
    | B^0\rangle}
    { \frac{8}{3}\left( \frac{f_B M_B}{Z_A(q^*)} \right)^2 }, 
\no
\ee
and operators are
given by
\be
        \hat{\Op^l}_L
&=&     \bar{b}\gamma_\mu(1-\gamma_5)d\
        \bar{b}\gamma_\mu(1-\gamma_5)d\no\\
        \hat{\Op^l}_R
&=&     \bar{b}\gamma_\mu(1+\gamma_5)d\
        \bar{b}\gamma_\mu(1+\gamma_5)d \no\\
        \hat{\Op^l}_N
&=&   2 \bar{b}\gamma_\mu(1-\gamma_5)d\
        \bar{b}\gamma_\mu(1+\gamma_5)d \no\\
& &+\ 4 \bar{b}(1-\gamma_5)d\
        \bar{b}(1+\gamma_5)d           \no\\
        \hat{\Op^l}_S
&=&     \bar{b}(1-\gamma_5)d\
        \bar{b}(1-\gamma_5)d.           \no
\ee
The operators other than $\hat{\Op^l}_L$ appear with
$O(\alpha_s)$ mixing coefficients.
Since the one-loop calculation of the coefficients $Z_i$ has not
yet been done for the NRQCD action, we use those in the static
approximation instead in this preliminary study.
$Z_i's$ at $\beta$=5.9  are given in Table \ref{matchfac}.

The method of the calculation of $B^{\rm lat}_i$'s is
standard.
We place a point source of the heavy and light quarks at the
origin, where the 4-fermi
\begin{table}[H]
\vspace{-0.8cm}
\begin{center}
\begin{tabular}{|c||c|c|c|c|c|}
\hline
$q^*$ & $Z_L^{\rm stat}$ & $Z_R^{\rm stat}$
      & $Z_N^{\rm stat}$ & $Z_S^{\rm stat}$
      & $Z_A^{\rm stat}$\\
\hline\hline
$\pi/a$ & 0.938 & -0.007 & -0.080 & -0.139 & 0.864 \\
$1/a$   & 0.998 & -0.011 & -0.132 & -0.120 & 0.800 \\
\hline
\end{tabular}
\caption{Matching factors at $\beta$=5.9.}
\label{matchfac}
\end{center}
\end{table}
\hspace{-0.5cm}
operators are constructed, and $B$
and $\bar{B}$ mesons propagate in the opposite direction to
each other.
The ground state extraction is rather easier for NRQCD than
for the static approximation, even though we use the local
sink.
This is another advantage of introducing the $1/m_Q$
corrections.

The actions and lattice parameters are summarized in Table
\ref{simpara}.
The inverse lattice spacing is determined from the rho meson
mass as $a^{-1}$ = 1.62 GeV.
For the heavy quark, we performed two sets of calculations
with $1/m_Q$ and $1/m_Q^2$ NRQCD actions as in Ref.\cite{our},
in order to estimate the higher order effects in the $1/m_Q$
expansion.
\begin{table}[t]
\begin{tabular}{|c|l|}
\hline
Size        & $16^3 \times 48\ \ (t = [ -24, 23 ])$\\
\hline
Gauge       & Wilson plaquette action\\
            & quenched 100 config.\\
            & $\beta$=5.9,\ fixed to Coulomb gauge\\
\hline
Light quark & SW-Clover action($c_{\rm sw}=1/u_0^3$)\\
            & local source at $t = 0$\\
            & KLM norm.\cite{KLM}\\
\hline
Heavy quark & Tadpole improved NRQCD\\
            & $O(1/m)$ and $O(1/m^2)$\cite{our}\\
            & local source at $t = 0$\\
\hline
\end{tabular}
\label{simpara}
\caption{Actions and simulation parameters}
\vspace{-0.8cm}
\end{table}
\section{Results}

\begin{center}
\begin{figure}
\leavevmode\psfig{file=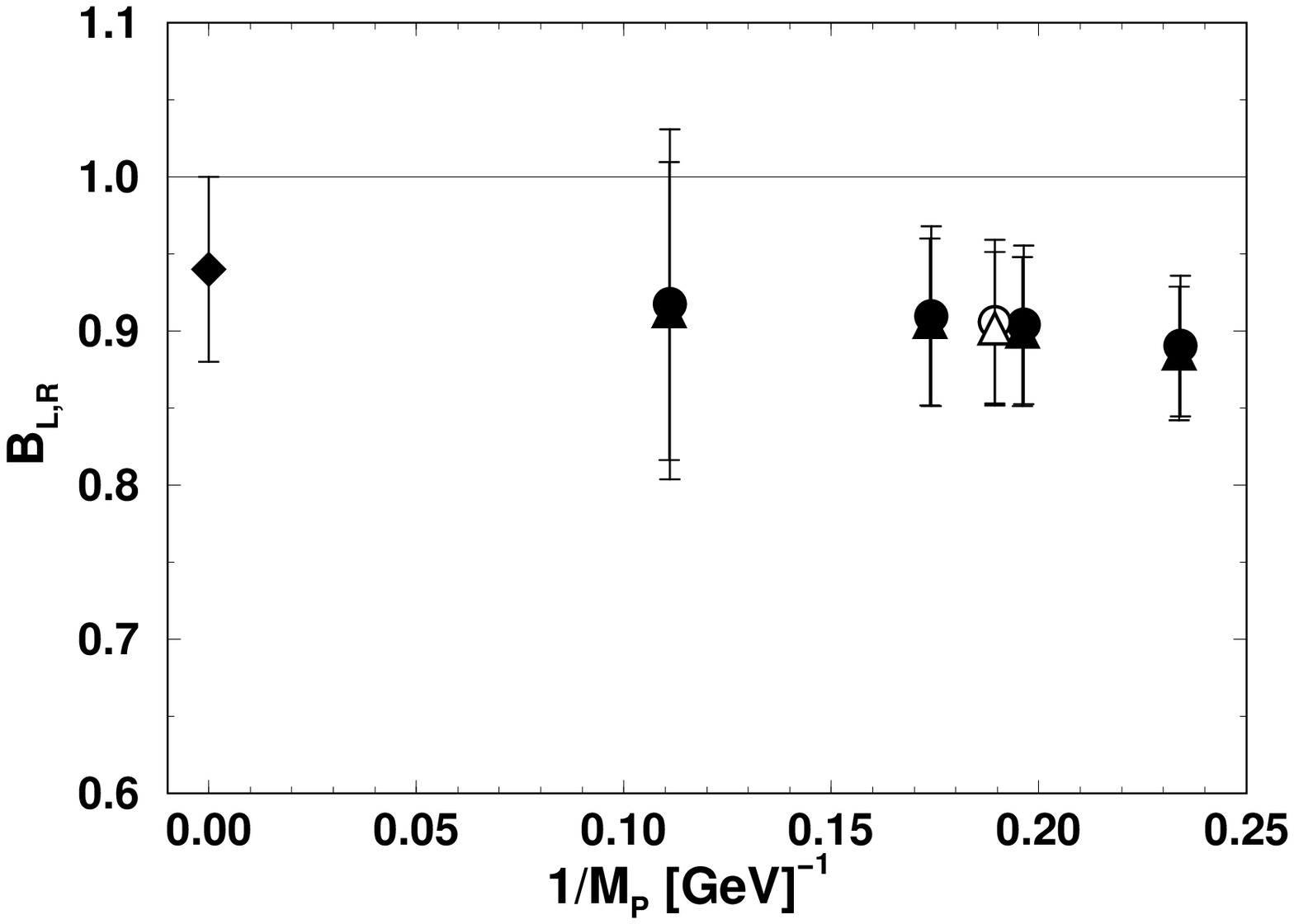,width=6.9cm}
\leavevmode\psfig{file=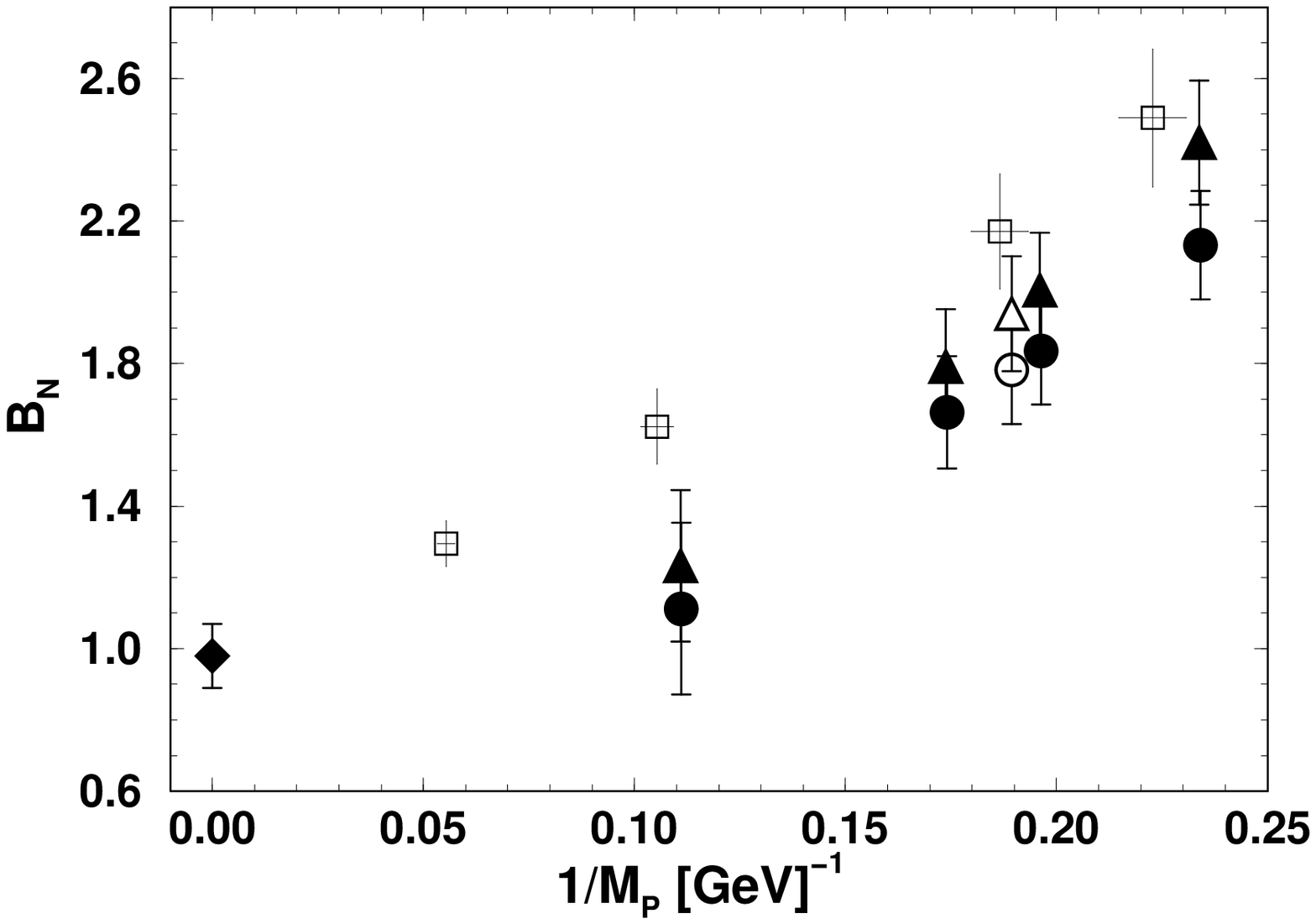,width=6.9cm}
\leavevmode\psfig{file=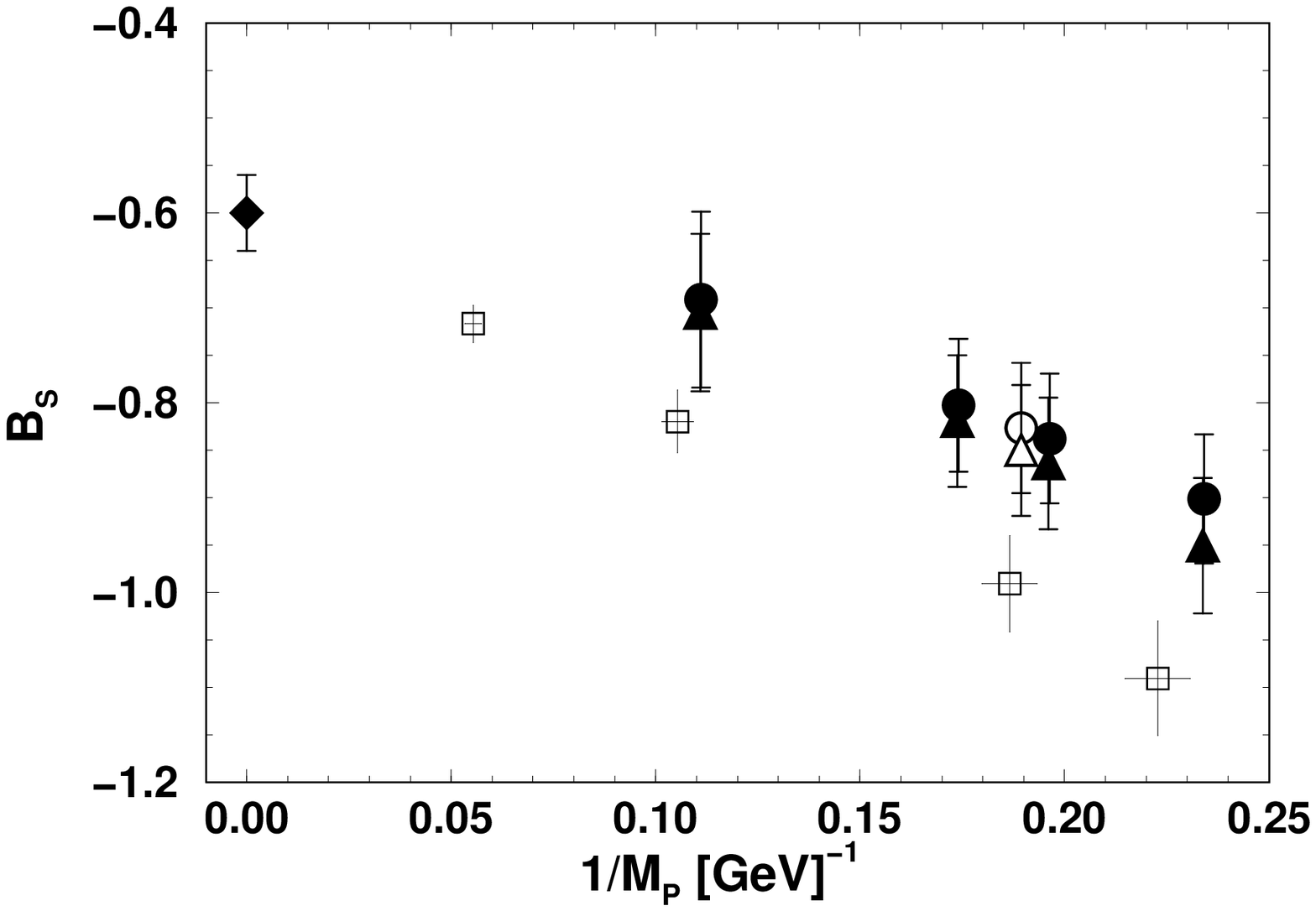,width=6.9cm}
\caption{The heavy quark mass dependence of $B_i^{\rm lat}$.
The circles are for $O(1/m_Q)$ calculation and
the triangles for $O(1/m_Q^2)$.
Opaque symbols are obtained by interpolations
to the physical $B$ meson mass, respectively.
Diamonds are the results obtained in Ref.\cite{GM}.}
\label{fig2}
\end{figure}
\end{center}

In Figure \ref{fig2} we plot the $1/M_P$ dependence of
$B^{\rm lat}_i$'s, with $M_P$ the pseudoscalar heavy-light
meson mass.
Since $B_L^{\rm lat}$ should agree with $B_R^{\rm lat}$
with infinite statistics, we take an average of
$B_L^{\rm lat}$ and $B_R^{\rm lat}$ in our simulation.
$B_N$ and $B_S$ have significant slopes in $1/M_P$, while
$B_{L,R}$ are almost flat.
The curvature in $1/M_P$, on the other hand, is small in any
case, and the difference between the two different NRQCD
actions is negligible.
An extrapolation to the static limit does not seem to agree
with the result of \cite{GR} (at $\beta$=6.0) for $B_N$ and
for $B_S$, although it is not quite clear whether this is an
inconsistency, with our large statistical errors.

\vspace{-0.8cm}
This $1/M_P$ dependence can be roughly understood by
the factorization hypothesis  (or vacuum saturation) as
follows.
If the factorization is exact on the lattice,
$B_{L,R}^{\rm (fac)}$ becomes unity by definition (solid
line in Figure \ref{fig2}).
For $B_{N,S}^{\rm (fac)}$, after a little algebra, we obtain
\be
      B_N^{\rm (fac)}
&=&   1 - 4 \frac{\delta f_P^{(1)}}{f_P}\label{eq2}\\
      B_S^{\rm (fac)}
&=& - \frac{5}{8}
      \left[ 1 - 8 \frac{\delta f_P^{(1)}}{f_P} \right]\label{eq3},
\ee
where $\delta f_P^{(1)}$ is the $O(1/m_Q)$ correction to
the decay constant $f_P$.
$\delta f_B^{(1)}$ is known to be rather large, and as an
estimate we use our previous calculation with NRQCD
\cite{our}.
Substituting the value of $\delta f_B^{(1)}/f_B$
to Eq.(\ref{eq2}) and (\ref{eq3}),
we obtain open squares in Fig.\ref{fig2} for $B_{N,S}^{\rm(fac)}$.
The agreement between the full calculation and the
factorization is remarkable.
Not only their value itself, but also the slope in $1/m_Q$
seem to be well described with the factorization hypothesis.

\begin{figure}
\leavevmode\psfig{file=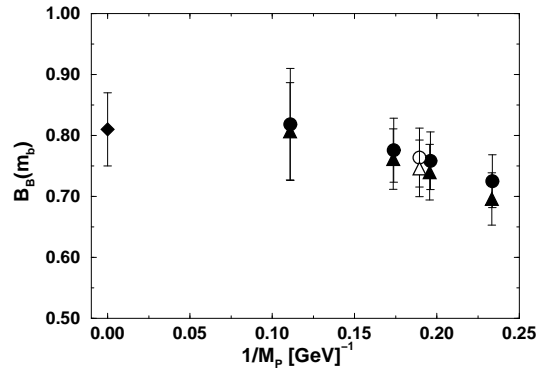,width=7cm}
\caption{The heavy quark mass dependence of $B_B(m_b)$.
A diamond is the result of reanalysis\cite{GR}.
For other symbols, see the previous figure caption.}
\label{fig3}
\end{figure}

The $1/M_P$ dependence of $B_B(m_b)$ is shown in Figure
\ref{fig3}, which shows a little negative slope.
This is because of a large cancellation of $1/M_P$ dependence
between those in $B_N$ and $B_S$.
In the leading log approximation,
with $\Lambda_{QCD}^{(5)}$=0.2 GeV,
we obtain the scale invariant $B_B$ parameter
\be
\hat{B}_{B_d} = (\alpha_s(m_b))^{-6/23} B_B(m_b) = 1.17(9)(4)(3)\no.
\ee
The errors are statistical, higher order in perturbation
theory, and the higher order in $1/m_Q$ expansion,
respectively.
For other quantities, we obtain
$B_{B_s}/B_{B_d} =  1.00(2)$ and $f_{B_s}/f_{B_d} = 1.16(3)$.
Some additional systematic uncertainties, such as
$O(a^2)$, $O(\alpha_s /m)$, $O(a \alpha_s)$ and quenching errors
still remain to be estimated.

\section{Summary}
To summarize,
we find a large $1/m_Q$ effects in the matrix elements
$B_N^{lat}$ and $B_S^{lat}$,
although the $O(1/m_Q^2)$ correction seems reasonably small. 
Their large mass dependence can be roughly understood
by the factorization hypothesis (or vacuum saturation).
In the present analysis, we combine lattice simulation
for finite heavy quark mass with matching coefficients
in the static limit, which is still unsatisfactory.
For the complete understanding of $1/m_Q$ dependence, 
matching coefficients with finite heavy quark mass are 
absolutely necessary.

\section{Acknowledgment}
Numerical calculations have been done on Paragon XP/S at
Institute for Numerical Simulations and Applied Mathematics
in Hiroshima University.
We are grateful to S. Hioki for allowing us to use his program
to generate gauge configurations.

\end{document}